\documentclass[prd,twocolumn,eqsecnum,nofootinbib]{revtex4}
\usepackage{amssymb}
\newcommand{\gothf}{\mathfrak{f}} 
\begin{document}
\title{Extended-body approach to the electromagnetic self-force in
  curved spacetime}  
\author{Javier Molina S\'anchez and Eric Poisson}
\affiliation{Department of Physics, University of Guelph, Guelph,
Ontario, Canada N1G 2W1}
\date{May 23, 2006; updated January 5, 2007}  
\begin{abstract}
We offer a novel derivation of the electromagnetic self-force acting
on a charged particle moving in an arbitrary curved spacetime. Our
derivation is based on a generalization from flat spacetime to curved
spacetime of the extended-body approach of Ori and Rosenthal. In this
approach the charged particle is first modeled as a body of finite
extension $s$, the net force acting on the extended body is computed,
and the limit $s \to 0$ is taken at the end of the
calculation. Concretely our extended body is a dumbbell that consists
of two point charges that are maintained at a constant spacelike
separation $s$. The net force acting on the dumbbell includes
contributions from the mutual forces exerted on each charge by the
field created by the other charge, the individual self-forces exerted
on each charge by its own field, and the external force which is
mostly responsible for the dumbbell's acceleration. These
contributions are added up, in a way that respects the curved nature  
of the spacetime, and all diverging terms in the net force are shown 
to be removable by mass renormalization. Our end result, in the limit
$s \to 0$, is the standard expression for the electromagnetic
self-force in curved spacetime. 
\end{abstract}
\maketitle

\section*{Disclaimer} 

An earlier version of this paper was submitted for publication in 
{\sl Physical Review D}, and it was severely criticised by a
knowledgeable referee. After revising the paper to fully acknowledge
the flaws of this work, and failing once more to convince the referee, 
we decided to withdraw the paper and not to pursue its publication. We
tried to do what should not be done: To derive the equations of motion
of a charged, extended body without formulating a consistent model for
its internal dynamics. This effort was misguided, and the referee was
correct to recommend against publication.  

Nevertheless, there is much that we like about this paper. First, we
believe that Fermi normal coordinates (Sec.\ II) are a very useful and 
natural tool for this problem. Second, we find it remarkable that the
total force is largely insensitive to the details of the rule (Sec.\
V) that is used to transport individual forces to a common point. And
finally, we believe that our critical review of the literature (Sec.\
VI) is valuable, and we reiterate our view that ``the existing
derivations [of the standard expression for the electromagnetic
self-force], including the one presented here, possess varying
strengths and weaknesses, but that none of them can be considered to
be completely satisfactory.'' We hope that this contribution will
motivate other researchers to seek a derivation that is fully beyond
reproach.   

\section{Introduction and overview}  

A particle with an electric charge $e$, subjected to an external
force $f_{\rm ext}^\mu$, moves in a curved spacetime according to the 
equations of motion 
\begin{equation} 
m a^\mu = f_{\rm ext}^\mu + f_{\rm self}^\mu, 
\label{1.1}
\end{equation} 
where $m$ is the particle's mass, $a^\mu = D u^\mu/d\tau$ is the
particle's covariant acceleration, the covariant derivative with
respect to proper time $\tau$ of the velocity vector $u^\mu$, and
$f_{\rm self}^\mu$ is the particle's electromagnetic self-force,
the force exerted on the particle by its own electromagnetic
field. This is given by  
\begin{eqnarray} 
\hspace*{-20pt} 
f_{\rm self}^\mu &=& 
e^2 \bigl( \delta^\mu_{\ \nu} + u^\mu u_\nu \bigr)  
\biggl( \frac{2}{3} \frac{D a^\nu}{d \tau}    
+ \frac{1}{3} R^{\nu}_{\ \lambda} u^{\lambda} \biggr)  
\nonumber \\ & & \mbox{} 
+ 2 e^2 u_\nu \int_{-\infty}^{\tau^-}     
\nabla^{[\mu} G^{\nu]}_{\ \lambda'}\bigl(z(\tau),z(\tau')\bigr)   
u^{\lambda'}\, d\tau'. 
\label{1.2}
\end{eqnarray} 
The self-force is expressed in terms of the velocity vector $u^\mu$,
the rate of change of the acceleration vector $Da^\mu/d\tau$, the
Ricci tensor $R^\mu_{\ \nu}$ of the curved spacetime, and in terms of
an integral over the particle's past history. The integration involves
the retarded Green's function $G^\mu_{\ \nu'}(x,x')$ associated with
the wave equation satisfied by the electromagnetic vector potential
(in the Lorenz gauge); the first argument of the Green's function is
the point $z(\tau)$, the current position of the particle, and its
second argument is a prior position $z(\tau')$. The integration also
involves the charge's velocity vector $u^{\mu'}$ at the prior position
$z(\tau')$. The integration is cut short at $\tau' = \tau^- \equiv
\tau - 0^+$ to avoid the singular behavior of the Green's function
when $z(\tau')$ coincides with $z(\tau)$. We shall refer to the
right-hand side of Eq.~(\ref{1.2}) as the {\it standard expression}
for the electromagnetic self-force. 

The standard expression was first obtained by DeWitt and Brehme
\cite{dewitt-brehme:60}, who based their derivation on
energy-momentum conservation, and who generalized Dirac's pioneering
analysis \cite{dirac:38} from flat spacetime to curved spacetime;
their calculations were later corrected by Hobbs \cite{hobbs:68}. The
standard expression was also derived on the basis of a logically
independent axiomatic approach by Quinn and Wald
\cite{quinn-wald:97}. Another alternative derivation was produced by
Detweiler and Whiting \cite{detweiler-whiting:03}, on the basis of
their unique decomposition of the retarded field into singular and
regular fields, and an assertion that the singular field exerts no
force on the particle. Finally, the standard expression was derived by
Poisson \cite{poisson:04b} on the basis of an averaging procedure,
wherein the self-force is assumed to be produced by the average of the
particle's retarded field over a spherical surface surrounding the
charge.   

Our purpose in this paper is to propose yet another derivation of the  
standard expression, and to critically review the derivations listed 
in the preceding paragraph. Our contention is that the existing
derivations, including the one presented here, possess varying
strengths and weaknesses, but that none of them can be considered to
be completely satisfactory. The derivation presented here is certainly
not immune to criticism. We believe, however, that its flaws are 
sufficiently different from the flaws of the competition that it
merits to be developed. Furthermore, we believe that the calculational
techniques introduced here will provide a useful basis for a future
attempt at providing a fully satisfactory derivation of the standard
expression.  

We have in mind an approach in which the charged object is
fundamentally not a point particle, but a body of finite extension
(denoted $s$) that is held together against electrostatic repulsion by
cohesive forces. A physical model for such an object would involve the
specification of its internal composition, a description of the
internal dynamics (the motion of each body element around the center 
of mass, in response to the applied cohesive and electromagnetic 
forces), and a derivation of its external dynamics (the motion of the 
center of mass in response to the external and self forces). This
modeling would have to be done in a fully relativistic setting that
takes into account the retardation of all interactions within the
body, and the curved nature of the background spacetime. In the limit
in which the length and time scales associated with the external
electromagnetic field, and those associated with the curved spacetime,
are large compared with $s$, we would expect this detailed calculation
to reproduce the standard expression of Eq.~(\ref{1.2}). The reason is
that in this limit, the external fields do not probe the internal
structure of the charged body, and the equations of motion would have
to reflect a near-complete decoupling of the internal and external
dynamics; this is achieved by the standard expression.  

The program outlined in the previous paragraph is a difficult one to
pursue. This approach has been very successful in the context of the
self-forced motion of a {\it massive body} (as opposed to an
electrically charged body) in a curved spacetime \cite{mino-etal:97, 
poisson:04a, poisson:04b}. Here the body could be modeled as a black
hole, an extended object with the simplest structure compatible with
the laws of general relativity. The motion of the black hole could be 
derived by matching the metric of the black hole perturbed by the
tidal gravitational field supplied by the external universe to the
metric of the external universe perturbed by the moving black
hole. Demanding that this metric be a valid solution to the Einstein
field equations determines the motion without additional input. The
end result is an equation of motion that makes no reference to the 
internal structure of the extended body. While the gravitational
implementation of this program was made possible by the simplicity
afforded by the black hole, an electromagnetic implementation would
have to involve a detailed modeling of the body's internal
structure. For this reason the electromagnetic case is much more
difficult to deal with. To the best of our knowledge, the
electromagnetic program has yet to be completed.   

In this paper we take some steps toward an eventual completion of this   
program. We consider a charged body with a very simple internal 
structure, but we make no attempt to model its internal dynamics. This    
is hardly satisfactory, and we are indeed a long way from reaching the
stated goals, but we nevertheless shall see that some measure of
success can be achieved with this minimalistic setting. In particular,
we shall see that our approach does permit a derivation of the
standard expression of Eq.~(\ref{1.2}). While we concede that our
approach is flawed and does not produce a fully satisfactory
derivation of Eq.~(\ref{1.2}), we believe that our contribution is
still valuable. In particular, as was stated above, we believe that
the calculational techniques introduced in this paper will be useful 
in an eventual completion of the program.        

Concretely our derivation is based on the Ori-Rosenthal extended-body 
approach to the electromagnetic self-force \cite{ori-rosenthal:03,
ori-rosenthal:04}, which we generalize from flat spacetime to curved
spacetime. Our extended body is a dumbbell that consists of two point   
charges that are maintained at a constant spacelike separation $s$;
the orientation of the dumbbell, relative to the direction of the
motion, is arbitrary. The dumbbell has no internal dynamics: the
separation between the two charges is kept constant by a method that
will be described in Sec.~IV. Because our extended body does not have
a continuous distribution of charge, our calculation of the net force
acting on it cannot proceed without an input axiom; this must provide
us with some information on the individual self-forces that act on the  
constituting charges. We want, of course, this axiom to be as mild as
possible, and we want the calculation to be capable of producing, in
the limit $s\to 0$, a precise and well-defined expression for the
electromagnetic self-force.             

Our input axiom is mild. It states:
\begin{verse} 
The self-force acting on a point charge is a well-defined vector field 
on the particle's world line, and it is proportional to the square of
the particle's electric charge.
\end{verse} 
This axiom is, first of all, an assumption that there exists a
self-force acting on each constituting point charge in the
dumbbell. This assumption is required by physical consistency: It
would be inconsistent to assume that the individual self-forces do not
exist, and to find in the end that a nonzero self-force emerges from 
the extended body in the limit $s\to 0$. The axiom is also an
assumption that the self-force is proportional to the square of 
the particle's charge. This property also follows from a requirement 
of physical consistency: In a spacetime with a timelike Killing
vector, the total work done by the self-force must equal the total
energy radiated by the charge \cite{quinn-wald:99}; because the
radiated energy is proportional to the square of the charge, we
require the same of the self-force.   

The formulation of the axiom constitutes the first flaw of our
minimalistic approach: such an axiom would not be required in an 
extended-body calculation that would deal honestly with the body's
internal dynamics. Moreover, the assumed equality between radiated 
energy and work done by the self-force must be only approximately
valid, because heat dissipation within an extended body would also
contribute to the energy balance. This effect, however, should become
unimportant in the limit $s \to 0$, that is, in the limit in which the
internal dynamics decouple from the external dynamics.     

Our derivation of the standard expression rests on the axiom, on
additional contrivances that will be introduced below, and on 
various computational devices that can be imported from the
literature. The geometry of our dumbbell is well adapted to a
description in terms of Fermi normal coordinates, and we make heavy
use of this coordinate system throughout the paper. The coordinates
are introduced in most textbooks on general relativity (see, for
example, Refs.~\cite{MTW:73} or \cite{poisson:b04}), but here we rely
on the presentation given in Poisson's review article, published
online in {\it Living Reviews in Relativity} \cite{poisson:04b}. We
will make heavy use of this resource in this paper; below it will be
repeatedly referred to as LRR. The properties of the Fermi normal
coordinates that are needed in our derivation are summarized in
Sec.~II.  

Another set of results that can be imported from the literature is 
the expression for the retarded electromagnetic field created by a 
point charge moving on an accelerated world line in a curved
spacetime. This field produces a force on a neighboring charge, and
this force contributes to the net force acting on the
dumbbell. Expressions for the field, given as expansions in
powers of $s$ in Fermi normal coordinates, are derived in LRR 
\cite{poisson:04b} and used here; these are presented in Sec.~III.  

A useful computational device for our derivation is the
Detweiler-Whiting decomposition \cite{detweiler-whiting:03} of the
retarded electromagnetic field $F_{\alpha\beta}$ into singular ``S''
and regular ``R'' pieces,   
\begin{equation} 
F_{\alpha\beta} = F^{\rm S}_{\alpha\beta} + F^{\rm R}_{\alpha\beta}.  
\label{1.3}
\end{equation} 
As was shown by Detweiler and Whiting, this decomposition is
unambiguous, the retarded and singular fields share the same
singularity structure near the world line, and the regular field  
$F^{\rm R}_{\alpha\beta}$ is smooth on the world line. They showed 
also that the retarded and singular fields satisfy the same field
equations (with a distributional current density on the right-hand
side), but that the regular field is sourcefree. The Detweiler-Whiting  
decomposition is introduced in Sec.~III, and it is used here without
making assumptions about the physical role of the singular field. We
involve it mostly as a matter of convenience, but also to emphasize
the point that the standard expression for the self-force describes an 
interaction between the particle and the regular field.           

The dumbbell is formally introduced in Sec.~IV, and in Sec.~V the net  
force acting on it is computed from the individual
contributions. These include the mutual forces exerted on each charge  
by the field created by the other charge, the individual self-forces
exerted on each charge by its own field, and the external force which
is mostly responsible for the dumbbell's acceleration. This step
requires another computational device, a transport rule that allows us 
to bring the different forces to a common point before adding them
up. Our transport rule is introduced in Sec.~V, and it is a
straightforward generalization, from flat spacetime to curved
spacetime, of the Ori-Rosenthal rule \cite{ori-rosenthal:03,
ori-rosenthal:04}.    

The final step of the calculation is to renormalize the mass and then 
take the limit $s\to 0$. These calculations also are presented in
Sec.~V, and the final outcome is Eq.~(\ref{1.2}), the standard
expression for the electromagnetic self-force. We conclude the paper
in Sec.~VI with a critical review of the existing derivations of the
self-force, in which we comment on their strengths and weaknesses. We
also offer a critical summary of our own approach.        

Throughout the paper we use the notations and conventions of Misner, 
Thorne, and Wheeler \cite{MTW:73}. In particular, we use geometrized
units in which $G = c = 1$.  

\section{Fermi coordinates and comoving world lines} 

\subsection{Fermi normal coordinates} 

We consider a neighborhood ${\cal N}$ around a reference world line  
$\bar{\gamma}$ in a curved spacetime with metric $g_{\alpha\beta}$;
the extension of ${\cal N}$ is small compared with the radius of
curvature of the spacetime. The world line is arbitrary, and it
possesses an acceleration. The neighborhood ${\cal N}$ is charted with
Fermi normal coordinates $(t,x^a)$, in which $t$ is proper time on the
world line, and $x^a$ ($a = 1, 2, 3$) are spatial coordinates that
vanish on the world line [LRR Sec.~3.2]. The coordinates have the
property that a spacelike geodesic orthogonal to $\bar{\gamma}$ is
described by the equations $t = \mbox{constant}$ and $x^a = s
\omega^a$, where $s$ is proper distance away from the world line
(measured along the geodesic), and the coefficients $\omega^a$ are
constant. They also have the property that $s^2 = \delta_{ab} x^a
x^b$, where $\delta_{ab}$ is the Kronecker delta, so that the
$\omega^a$'s form the components of a unit vector, $\delta_{ab}
\omega^a \omega^b = 1$.      

The metric of the curved spacetime in ${\cal N}$ is expressed in Fermi
normal coordinates (FNC) as an expansion in powers of the spatial
coordinates $x^a$. It takes the form [LRR Eq.~(125)--(127)] 
\begin{eqnarray} 
\hspace*{-15pt} 
g_{tt} &=& -1 - 2\bar{a}_a x^a - (\bar{a}_a x^a)^2 
- \bar{R}_{ctdt} x^c x^d + O(s^3),   
\label{2.1} \\ 
\hspace*{-15pt} 
g_{ta} &=& -\frac{2}{3} \bar{R}_{actd} x^c x^d + O(s^3), 
\label{2.2} \\ 
\hspace*{-15pt} 
g_{ab} &=& \delta_{ab} 
- \frac{1}{3} \bar{R}_{acbd} x^c x^d + O(s^3), 
\label{2.3}
\end{eqnarray} 
where $\bar{a}_a(t) \equiv \delta_{ab} \bar{a}^b(t)$ are the spatial
components of the vector $a_{\bar{\alpha}}$, $\bar{\gamma}$'s
acceleration vector, in FNC (the time component vanishes), and where
$\bar{R}_{ctdt}(t)$, $\bar{R}_{actd}(t)$, and $\bar{R}_{acbd}(t)$ are 
the components of the Riemann tensor evaluated on $\bar{\gamma}$;
these depend on $t$ only. Here and below we use an overbar to
indicate quantities that are evaluated on $\bar{\gamma}$. When the
quantity is the component of a tensor in FNC, we place the overbar on
the tensorial symbol, as in $\bar{a}_a$; when the quantity is the
tensor itself, we place the overbar on the tensorial index, as in 
$a_{\bar{\alpha}}$. 

The FNC have the property that the metric reduces to Minkowski
values on the reference world line. The inverse metric is  
\begin{eqnarray} 
g^{tt} &=& -1 + 2 \bar{a}_a x^a - 3(\bar{a}_a x^a)^2 
+ \bar{R}_{ctdt} x^c x^d 
\nonumber \\ & & \mbox{} 
+ O(s^3), 
\label{2.4} \\ 
g^{ta} &=& -\frac{2}{3} \bar{R}^a_{\ ctd} x^c x^d + O(s^3), 
\label{2.5} \\  
g^{ab} &=& \delta^{ab} 
+ \frac{1}{3} \bar{R}^{a\ b}_{\ c\ d} x^c x^d + O(s^3). 
\label{2.6} 
\end{eqnarray} 
Here the indices on the Riemann tensor are raised freely with
$\delta^{ab}$; this is appropriate because the Riemann tensor is 
evaluated on the reference world line. 

It is useful to decompose the metric in terms of a tetrad of dual
vectors $e^{(0)}_\alpha$, $e^{(a)}_\alpha$ that are orthonormal
everywhere in ${\cal N}$. Here the superscript within brackets
serves to label each one of the four tetrad members, and the Greek 
subscript is the usual vectorial index.\footnote{The dual tetrad is
denoted $\bar{e}^0_\alpha$, $\bar{e}^a_\alpha$ in LRR.} 
The decomposition is given by  
\begin{equation} 
g_{\alpha\beta} = -e^{(0)}_\alpha e^{(0)}_\beta 
+ \delta_{ab} e^{(a)}_\alpha e^{(b)}_\beta 
\label{2.7}
\end{equation}
and the members of the tetrad are [LRR Eqs.~(123) and (124)] 
\begin{eqnarray}
e^{(0)}_t &=& 1 + \bar{a}_a x^a 
+ \frac{1}{2} \bar{R}_{ctdt} x^c x^d + O(s^3), 
\label{2.8} \\ 
e^{(0)}_a &=& \frac{1}{6} \bar{R}_{actd} x^c x^d + O(s^3), 
\label{2.9} \\ 
e^{(a)}_t &=& -\frac{1}{2} \bar{R}^a_{\ ctd} x^c x^d + O(s^3), 
\label{2.10} \\ 
e^{(a)}_b &=& \delta^a_{\ b} 
- \frac{1}{6} \bar{R}^a_{\ cbd} x^c x^d + O(s^3). 
\label{2.11}
\end{eqnarray} 
The tetrad vectors possess the important property that they are
parallel transported along the spacelike geodesic that leaves
$\bar{\gamma}$ orthogonally to arrive at the point $x = (t,x^a)$. As
was mentioned previously, this spacelike geodesic is described by the
equations $t = \mbox{constant}$ and $x^a = s \omega^a$, with constant 
coefficients $\omega^a$; its departure point on $\bar{\gamma}$ is
therefore $\bar{x} = (t,0)$.  

We introduce also a tetrad of vectors $e^\alpha_{(0)}$, 
$e^\alpha_{(a)}$ which is related to the dual tetrad by   
\begin{equation} 
e^\alpha_{(0)} = -g^{\alpha\beta} e^{(0)}_{\beta}, \qquad 
e^\alpha_{(a)} = \delta_{ab} g^{\alpha\beta} e^{(b)}_{\beta}. 
\label{2.12}
\end{equation} 
With Eqs.~(\ref{2.4})--(\ref{2.11}) we obtain the explicit forms 
\begin{eqnarray} 
e^t_{(0)} &=& 1 - \bar{a}_a x^a + (\bar{a}_a x^a)^2 
- \frac{1}{2} \bar{R}_{ctdt} x^c x^d 
\nonumber \\ & & 
+ O(s^3), 
\label{2.13} \\ 
e^a_{(0)} &=& \frac{1}{2} \bar{R}^a_{\ ctd} x^c x^d + O(s^3),
\label{2.14} \\ 
e^t_{(a)} &=& -\frac{1}{6} \bar{R}_{actd} x^c x^d + O(s^3), 
\label{2.15} \\ 
e^b_{(a)} &=& \delta^b_{\ a} 
+ \frac{1}{6} \bar{R}^b_{\ cad} x^c x^d + O(s^3). 
\label{2.16}
\end{eqnarray} 
Any vector $A^\alpha$ at a point $x = (t,x^a)$ in ${\cal N}$ can be
decomposed in the tetrad $e^\alpha_{(0)}$, $e^\alpha_{(a)}$; the
decomposition takes the form of 
\begin{equation} 
A^\alpha = A^{(0)} e^\alpha_{(0)} + A^{(a)} e^\alpha_{(a)}, 
\label{2.17}
\end{equation} 
and the coefficients $A^{(0)} = A^\alpha e^{(0)}_\alpha$, $A^{(a)} =
A^\alpha e^{(a)}_\alpha$ are the frame components of the vector at the
point $x$.   

Any vector $A^\alpha$ at the point $x = (t,x^a)$ can be parallel 
transported to the simultaneous point $\bar{x} = (t,0)$ on
$\bar{\gamma}$ by following the spacelike geodesic $t =
\mbox{constant}$ that links these points. Because the tetrad vectors
are themselves parallel transported along this geodesic, the frame
components of the vector stay constant during the transport. The
vector at $\bar{x}$ is therefore $A^{\bar{\alpha}} 
= A^{(0)} e^{\bar{\alpha}}_{(0)} + A^{(a)} e^{\bar{\alpha}}_{(a)}$,
where $e^{\bar{\alpha}}_{(0)}$ and $e^{\bar{\alpha}}_{(a)}$ are the
tetrad vectors at $\bar{x}$. This can be expressed as 
\begin{equation} 
A^{\bar{\alpha}}(\bar{x}) = g^{\bar{\alpha}}_{\ \beta}(\bar{x},x)
A^{\beta}(x),    
\label{2.18} 
\end{equation} 
where $g^{\bar{\alpha}}_{\ \beta} = e^{\bar{\alpha}}_{(0)} 
e^{(0)}_{\beta} + e^{\bar{\alpha}}_{(a)} e^{(a)}_{\beta}$ is the
parallel propagator. Its components in FNC are
\begin{eqnarray} 
g^{\bar{t}}_{\ t} &=& 1 + \bar{a}_a x^a 
+ \frac{1}{2} \bar{R}_{ctdt} x^c x^d + O(s^3), 
\label{2.19} \\ 
g^{\bar{t}}_{\ a} &=& \frac{1}{6} \bar{R}_{actd} x^c x^d + O(s^3), 
\label{2.20} \\ 
g^{\bar{a}}_{\ t} &=& -\frac{1}{2} \bar{R}^a_{\ ctd} x^c x^d 
+ O(s^3),
\label{2.21} \\ 
g^{\bar{a}}_{\ b} &=& \delta^a_{\ b} 
- \frac{1}{6} \bar{R}^a_{\ cbd} x^c x^d + O(s^3). 
\label{2.22}
\end{eqnarray} 

\subsection{Congruence of comoving world lines} 

We introduce a congruence of comoving world lines $\gamma(x^a)$ in
${\cal N}$. These timelike curves are all described by $x^a =
\mbox{constant}$, and they are comoving with $\bar{\gamma} \equiv
\gamma(0)$ in the sense that they preserve their spatial coordinates,
and therefore their spatial separation $s$ relative to $\bar{\gamma}$,
during their motion. The comoving world lines are labeled by $x^a$ and
are parameterized by their proper time $\tau(x^a)$. This is related to
$t \equiv \bar{\tau} \equiv \tau(0)$ by $d\tau^2 = -g_{tt}\, dt^2$, so
that  
\begin{equation} 
\frac{d\tau}{dt} = 1 + \bar{a}_a x^a 
+ \frac{1}{2} \bar{R}_{ctdt} x^c x^d + O(s^3) 
\label{2.23}
\end{equation}
according to Eq.~(\ref{2.1}). The inverse relation is 
\begin{equation} 
\frac{dt}{d\tau} = 1 - \bar{a}_a x^a + (\bar{a}_a x^a)^2  
- \frac{1}{2} \bar{R}_{ctdt} x^c x^d + O(s^3). 
\label{2.24}
\end{equation}

The congruence's velocity field is denoted $u^\alpha$, and its 
components in FNC are 
\begin{equation} 
u^t = \frac{dt}{d\tau}, \qquad 
u^a = 0. 
\label{2.25}
\end{equation}
This vector field is everywhere tangent to the members of the
congruence. It is decomposed in the tetrad of 
Eqs.~(\ref{2.13})--(\ref{2.16}) as 
\begin{equation} 
u^\alpha = u^{(0)} e^\alpha_{(0)} + u^{(a)} e^\alpha_{(a)}, 
\label{2.26}
\end{equation} 
and the frame components are calculated to be 
\begin{equation} 
u^{(0)} = 1 + O(s^3), \qquad 
u^{(a)} = -\frac{1}{2} \bar{R}^a_{\ ctd} x^c x^d + O(s^3). 
\label{2.27}
\end{equation} 

The acceleration field of the congruence is defined as $a^\alpha =
u^\beta \nabla_\beta u^\alpha$; this is the acceleration vector of
each comoving world line. A straightforward calculation, based on the
Christoffel symbols obtained from Eqs.~(\ref{2.1})--(\ref{2.3}),
produces its components in FNC. We obtain 
\begin{equation} 
a^t = O(s^2), \qquad 
a^a = (1 - \bar{a}_c x^c) \bar{a}^a + \bar{R}^a_{\ tct} x^c + O(s^2); 
\label{2.28}
\end{equation} 
as expected, $a^t = 0$ and $a^a = \bar{a}^a$ when $x^c = 0$. The
acceleration field also can be decomposed as 
\begin{equation} 
a^\alpha = a^{(0)} e^\alpha_{(0)} + a^{(a)} e^\alpha_{(a)}. 
\label{2.29}
\end{equation} 
Its frame components are 
\begin{equation} 
a^{(0)} = O(s^2), \qquad 
a^{(a)} = (1 - \bar{a}_c x^c) \bar{a}^a + \bar{R}^a_{\ tct} x^c 
+ O(s^2). 
\label{2.30}
\end{equation} 

We conclude this section by introducing one additional vector field in
${\cal N}$. Take the spatial coordinates $x^a$ of the point $x$ and
divide by the proper distance $s = (\delta_{ab} x^a x^b)^{1/2}$
between $x$ and $\bar{x}$; this gives $\omega^a = x^a/s$, and these
form the components of a unit vector. Then define the vector field  
\begin{equation} 
\omega^\alpha = \omega^a e^\alpha_{(a)}. 
\label{2.31}
\end{equation} 
This is well-defined provided that $x^a \neq 0$; the vector field is
singular on $\bar{\gamma}$ by virtue of its directional ambiguity. It
is easy to see that this vector is normalized, in the sense that
$g_{\alpha\beta} \omega^\alpha \omega^\beta = 1$. Its most important
property is that it is tangent to the spacelike geodesic that links
the points $x$ and $\bar{x}$. This can be established from the
definition of the FNC in terms of Synge's world function
$\sigma(x,\bar{x})$ [LRR Secs.~2.1 and 3.2, especially Eq.~(116)]: We
have $x^a = -e^{(a)}_{\bar{\alpha}} \sigma^{\bar{\alpha}}$, where 
$e^{(a)}_{\bar{\alpha}}$ is the dual tetrad at $\bar{x}$ and 
$\sigma^{\bar{\alpha}} = \nabla^{\bar{\alpha}} \sigma(x,\bar{x})$, and
it follows that 
\[
\omega^\alpha = -\frac{1}{s} e^\alpha_{(a)} e^{(a)}_{\bar{\alpha}}  
\sigma^{\bar{\alpha}} = -\frac{1}{s} 
g^\alpha_{\ \bar{\alpha}}(x,\bar{x}) \sigma^{\bar{\alpha}}, 
\]
where we have involved the definition of the parallel propagator that
follows Eq.~(\ref{2.18}) and used the fact that
$e^{(0)}_{\bar{\alpha}} \sigma^{\bar{\alpha}} = -u_{\bar{\alpha}} 
\sigma^{\bar{\alpha}} = 0$. Using now the identity  
$-g^\alpha_{\ \bar{\alpha}} \sigma^{\bar{\alpha}} = \sigma^\alpha$,
where $\sigma^\alpha = \nabla^\alpha \sigma(x,\bar{x})$, we arrive at  
\begin{equation} 
\omega^\alpha = \frac{1}{s} \sigma^{\alpha}. 
\label{2.32}
\end{equation} 
The vector $\sigma^\alpha/s$ is normalized and it does have the
property of being tangent to the geodesic linking $x$ to $\bar{x}$
[LRR Sec.~2.1]. We note that the vector of Eq.~(\ref{2.32}) is defined
in a covariant way; the spatial distance $s$ is an invariant related
to Synge's world function by $s = \sqrt{2\sigma(x,\bar{x})}$.     

\section{Force on a comoving charge exerted by another comoving
  charge}  

We place an electric charge $\bar{e}$ on the reference world line
$\bar{\gamma}$, and we let it create a retarded electromagnetic field
$F_{\alpha\beta}$. We next place an electric charge $e$ on the world
line $\gamma \equiv \gamma(s\omega^a)$, a specific member of the
congruence introduced in Sec.~II B. Our specific world line is labeled
by its Fermi normal coordinates (FNC) $x^a = s \omega^a$, where $s$ is
$\gamma$'s constant spatial separation relative to $\bar{\gamma}$, and
$\omega^a$ are constant directional coefficients. The electromagnetic
field created by $\bar{e}$ exerts a force on $e$, and this is given by  
\begin{equation} 
f^\alpha = e F^{\alpha}_{\ \beta} u^\beta 
= f^{(0)} e^\alpha_{(0)} + f^{(a)} e^\alpha_{(a)}. 
\label{3.1}
\end{equation} 
Here $u^\alpha$ is the velocity vector of the charge $e$ moving on the
world line $\gamma$, whose components are listed in Eqs.~(\ref{2.24})
and (\ref{2.25}). In the second form the force is decomposed in the
tetrad of Eqs.~(\ref{2.13})--(\ref{2.16}). In these expressions we
must substitute $x^a = s \omega^a$ to select our particular member
$\gamma$ from the congruence $\gamma(x^a)$. 

If we further decompose $F_{\alpha\beta}$ and $u^\alpha$ in the tetrad
$e^\alpha_{(0)}$, $e^\alpha_{(a)}$ we find that the frame components
of the force are given by 
\begin{equation}
f^{(0)} = e F_{(a)(0)} u^{(a)}, \qquad 
f^{(a)} = e \bigl( F^{(a)}_{\ \ (0)} u^{(0)} 
+ F^{(a)}_{\ \ (b)} u^{(b)} \bigr). 
\label{3.2}
\end{equation} 
The frame components of the electromagnetic field were computed, and
expressed as expansions in powers of $s$ in FNC, in [LRR Eqs.~(455)
and (456)]. They are given by\footnote{The frame components are
denoted $\bar{F}_{a0}$ and $\bar{F}_{ab}$ in LRR, and contrary to our
usage here, quantities that refer to the reference world line
$\bar{\gamma}$ are presented in LRR without an overbar.}  
\begin{eqnarray} 
F_{(a)(0)} &=& \bar{e} \biggl[ 
\frac{1}{s^2} \omega_a 
- \frac{1}{2s} \bigl( \bar{a}_a + \bar{a}_b \omega^b \omega_a \bigr)  
+ \frac{3}{4} \bar{a}_b \omega^b \bar{a}_a 
\nonumber \\ & & \mbox{} 
+ \frac{3}{8} \bigl( \bar{a}_b \omega^b \bigr)^2 \omega_a 
+ \frac{3}{8} \dot{\bar{a}}_t \omega_a
- \frac{2}{3} \bar{R}_{atbt} \omega^b 
\nonumber \\ & & \mbox{} 
- \frac{1}{6} \bar{R}_{btct} \omega^b \omega^c \omega_a 
\nonumber \\ & & \mbox{} 
+ \frac{1}{12} \bigl( 5 \bar{R}_{tt} + \bar{R}_{bc} \omega^b\omega^c 
                      + \bar{R} \bigr) \omega_a 
\nonumber \\ & & \mbox{} 
- \frac{1}{6} \bar{R}_{ab} \omega^b \biggr] 
+ F_{(a)(0)}^{\rm R} + O(s) 
\label{3.3}
\end{eqnarray} 
and 
\begin{eqnarray} 
\hspace*{-15pt}
F_{(a)(b)} &=& \bar{e} \biggl[ 
\frac{1}{2} \bigl( \omega_a \dot{\bar{a}}_b 
                   - \dot{\bar{a}}_a \omega_b \bigr)   
+ \frac{1}{2} \bigl( \bar{R}_{atbc} - \bar{R}_{btac} \bigr) \omega^c  
\nonumber \\ & & \mbox{} 
- \frac{1}{2} \bigl( \bar{R}_{at} \omega_b 
                    - \omega_a \bar{R}_{bt} \bigr) \biggr] 
+ F_{(a)(b)}^{\rm R} + O(s). 
\label{3.4} 
\end{eqnarray} 
The quantities that appear in Eqs.~(\ref{3.3}) and (\ref{3.4}) were,
for the most part, introduced in Sec.~II. Exceptions are
$\dot{\bar{a}}_t$ and $\dot{\bar{a}}_a$, which are the components in
FNC of the vector $\dot{a}_{\bar{\alpha}} 
= D a_{\bar{\alpha}}/d\bar{\tau}$, the covariant derivative with
respect to proper time $\bar{\tau} \equiv \tau(0) \equiv t$ of
$\bar{\gamma}$'s acceleration vector. Additional exceptions are
$\bar{R}_{tt}$, $\bar{R}_{ta}$, $\bar{R}_{ab}$, and $\bar{R}$,
which are respectively the components in FNC of the Ricci tensor and  
the Ricci scalar, all evaluated on the reference world line
$\bar{\gamma}$. It should be noted that in Eqs.~(\ref{3.3}) and
(\ref{3.4}), $\omega_a \equiv \delta_{ab} \omega^b$. 

We have also introduced the frame components of the Detweiler-Whiting
regular field [Eq.~(\ref{1.3}); LRR Eqs.~(471) and (472)],  
\begin{equation} 
F_{(a)(0)}^{\rm R} = \bar{e} \biggl[ 
\frac{2}{3} \dot{\bar{a}}_a + \frac{1}{3} \bar{R}_{at} \biggr] 
+ \bar{F}_{(a)(0)}^{\rm tail} + O(s) 
\label{3.5}
\end{equation} 
and
\begin{equation} 
F_{(a)(b)}^{\rm R} = \bar{F}_{(a)(b)}^{\rm tail} + O(s). 
\label{3.6}
\end{equation} 
These are related to the frame components of the ``tail part'' of the  
field [LRR Eqs.~(453) and (454)],   
\begin{equation} 
\bar{F}_{(a)(0)}^{\rm tail} = 
F^{\rm tail}_{\bar{\alpha}\bar{\beta}} 
e^{\bar{\alpha}}_{(a)} e^{\bar{\beta}}_{(0)}, \qquad 
\bar{F}_{(a)(b)}^{\rm tail} =  
F^{\rm tail}_{\bar{\alpha}\bar{\beta}}
e^{\bar{\alpha}}_{(a)} e^{\bar{\beta}}_{(b)}, 
\label{3.7}
\end{equation} 
where 
\begin{equation} 
F_{\bar{\alpha}\bar{\beta}}^{\rm tail}(\bar{x}) 
= 2 e \int_{-\infty}^{\bar{\tau}^-} 
\nabla_{[\bar{\alpha}} G_{\bar{\beta}]\mu'}(\bar{x},z') 
u^{\mu'}\, d\bar{\tau}'.
\label{3.8}
\end{equation} 
Notice that the tail field is evaluated at the point $\bar{x} = (t,0)$
on the reference world line. It is expressed as an integral over
$\bar{e}$'s past history, for times ranging from $t' = -\infty$ to 
$t' = t^- \equiv t - 0^+$. The first argument of the Green's function
is the point $\bar{x}$, the current position of the charge $\bar{e}$,
and its second argument is a prior position $z' = (t',0)$. The tail
field is presented in a covariant form by identifying 
$\bar{\tau}' \equiv \tau'(0) \equiv t'$ and $\bar{\tau} \equiv t$.     

If we substitute Eqs.~(\ref{2.27}), (\ref{3.3}), and (\ref{3.4}) into
Eq.~(\ref{3.2}) we obtain the frame components of the force exerted on
the charge $e$. We find 
\begin{equation} 
f^{(0)} = O(s), \qquad 
f^{(a)} = e F^{(a)}_{\ (0)} + O(s). 
\label{3.9}
\end{equation}
We may now substitute these into Eq.~(\ref{3.1}) and reconstruct a
vectorial expression for the force. The vector $f^\alpha$ is at first
expressed in terms of the vectors $\omega^a e^{\alpha}_{(a)}$, 
$\bar{a}^a e^{\alpha}_{(a)}$, and quantities of the form 
$e^{\alpha}_{(a)} \bar{R}^a_{\ tct}\omega^c$ and $e^{\alpha}_{(a)}
\bar{R}^a_{\ b} \omega^b$. The vector field $\omega^\alpha = \omega^a
e^{\alpha}_{(a)}$ was already introduced in Eq.~(\ref{2.31}), and its
geometrical meaning was described at the end of Sec.~II. The vector
field $\bar{a}^a e^{\alpha}_{(a)}$ can easily be related to
$a^\alpha$, the acceleration field of Eq.~(\ref{2.28}); by combining 
Eqs.~(\ref{2.15}), (\ref{2.16}), and (\ref{2.28}) we obtain 
\begin{equation}
\bar{a}^a e^{\alpha}_{(a)} = (1 + s \bar{a}_c \omega^c) a^\alpha 
- s e^{\alpha}_{(a)} \bar{R}^a_{\ tct} \omega^c + O(s^2). 
\label{3.10}
\end{equation} 
Making this substitution into Eq.~(\ref{3.9}) and involving
Eq.~(\ref{3.3}), we obtain 
\begin{eqnarray} 
f^\alpha &=& e \bar{e} \biggl[  
\frac{1}{s^2} \omega^\alpha  
- \frac{1}{2s} a^\alpha 
- \frac{1}{2s} \bigl( \bar{a}_b \omega^b \bigr) \omega^\alpha 
+ \frac{1}{4} \bigl( \bar{a}_b \omega^b \bigr) a^\alpha  
\nonumber \\ & & \mbox{} 
+ \frac{3}{8} \bigl( \bar{a}_b \omega^b \bigr)^2 \omega^\alpha  
+ \frac{3}{8} \dot{\bar{a}}_t \omega^\alpha 
- \frac{1}{6} e^{\alpha}_{(a)} \bar{R}^a_{\ tbt} \omega^b 
\nonumber \\ & & \mbox{} 
- \frac{1}{6} \bar{R}_{btct} \omega^b \omega^c \omega^\alpha  
+ \frac{1}{12} \bigl( 5 \bar{R}_{tt} + \bar{R}_{bc} \omega^b\omega^c 
                      + \bar{R} \bigr) \omega^\alpha  
\nonumber \\ & & \mbox{} 
- \frac{1}{6} e^{\alpha}_{(a)} \bar{R}^a_{\ b} \omega^b \biggr] 
+ e e^{\alpha}_{(a)} F^{(a)}_{{\rm R} (0)} + O(s).  
\label{3.11} 
\end{eqnarray} 
This expression for the force involves the vectors $\omega^\alpha$,
$a^\alpha$, and $e^\alpha_{(a)}$ which are all appropriately defined
at $x = (t,s\omega^a)$, where the charge $e$ resides. But it also
involves quantities such as $\bar{a}_a$, $\dot{\bar{a}}_t$, 
$\bar{R}^a_{\ tct}$, etc., which are defined instead at $\bar{x}$. The
result is cumbersome, and our expression for the force needs to be
rationalized. We shall therefore endeavor to express $f^\alpha$
entirely in terms of quantities that are defined at $x$.  

To see how this can be achieved we present the details of a
representative piece of the calculation: We shall show that the
quantity $\bar{a}_a \omega^a$ that appears in Eq.~(\ref{3.11}) can be  
expressed as  
\begin{equation} 
\bar{a}_a \omega^a = a_\alpha \omega^\alpha 
+ s (a_\alpha \omega^\alpha)^2 
- s R_{\alpha\mu\beta\nu} \omega^\alpha u^\mu \omega^\beta u^\nu 
+ O(s^2), 
\label{3.12}
\end{equation}    
in terms of tensorial quantities defined at $x$ only. Here $a^\alpha$
is the acceleration field of Eq.~(\ref{2.28}), $u^\alpha$ is the
velocity field of Eqs.~(\ref{2.24}) and (\ref{2.25}), and 
$R_{\alpha\mu\beta\nu}$ is the Riemann tensor evaluated at $x$. 

We begin by stating that by virtue of Eq.~(\ref{2.31}) and the
orthonormality of the tetrad vectors, $\bar{a}^a \omega_a$ is equal to
$\bar{a}^a e^\alpha_{(a)} \omega_\alpha$. Combining this statement
with Eq.~(\ref{3.10}) gives 
\[
\bar{a}_a \omega^a = (1 + s \bar{a}_c \omega^c) a_\alpha \omega^\alpha  
- s \omega_\alpha e^{\alpha}_{(a)} \bar{R}^a_{\ tct} \omega^c 
+ O(s^2),  
\]
or 
\[
\bar{a}_a \omega^a = a_\alpha \omega^\alpha 
+ s \bigl( a_\alpha \omega^\alpha \bigr)^2   
- s \omega_\alpha e^{\alpha}_{(a)} \bar{R}^a_{\ tct} \omega^c 
+ O(s^2). 
\]
By involving the tetrad vectors of Eqs.~(\ref{2.8})--(\ref{2.16}), the 
components of the Riemann tensor in FNC can be expressed in covariant
form as  
\[
\bar{R}^a_{\ tct} 
= R^{\bar{\mu}}_{\ \bar{\lambda}\bar{\nu}\bar{\rho}} 
e^{(a)}_{\bar{\mu}} e^{\bar{\lambda}}_{(0)} e^{\bar{\nu}}_{(c)} 
e^{\bar{\rho}}_{(0)}. 
\]
The right-hand side is a scalar quantity, and this scalar can be
evaluated at $x$ instead of $\bar{x}$ at the cost of introducing a
correction term of order $s$. We then have  
\[
\bar{R}^a_{\ tct} 
= R^{\mu}_{\ \lambda\nu\rho} 
e^{(a)}_{\mu} e^{\lambda}_{(0)} e^{\nu}_{(c)} e^{\rho}_{(0)} + O(s), 
\]
and it follows that 
\[
e^{\alpha}_{(a)} \bar{R}^a_{\ tct} \omega^c = 
R^{\mu}_{\ \lambda\nu\rho} \bigl(e^{\alpha}_{(a)} e^{(a)}_{\mu} \bigr) 
e^{\lambda}_{(0)} \bigl( \omega^c e^{\nu}_{(c)} \bigr) e^{\rho}_{(0)} 
+ O(s).
\]
The factor within the first set of brackets is $\delta^\alpha_{\ \mu}
- e^\alpha_{(0)} e^{(0)}_{\mu}$, and the factor within the second set
is $\omega^\nu$; the right-hand side becomes
\[
R^{\alpha}_{\ \lambda\nu\rho} e^{\lambda}_{(0)} \omega^\nu
e^{\rho}_{(0)} + O(s).
\]
We now compare Eqs.~(\ref{2.13}), (\ref{2.14}), and (\ref{2.24}),
(\ref{2.25}) and observe that $e^\alpha_{(0)} = u^\alpha +
O(s^2)$. This allows us to write, finally, 
\[
e^{\alpha}_{(a)} \bar{R}^a_{\ tct} \omega^c = 
R^{\alpha}_{\ \lambda\nu\rho} u^\lambda \omega^\nu u^\rho + O(s), 
\]
and we have obtained Eq.~(\ref{3.12}). 

Working on each term of Eq.~(\ref{3.11}) in this fashion, we
eventually obtain our final expression 
\begin{eqnarray} 
f^\alpha &=& e \bar{e} \biggl[ 
\frac{1}{s^2} \omega^\alpha  
- \frac{1}{2s} a^\alpha 
- \frac{1}{2s} \bigl( a_\beta \omega^\beta \bigr) \omega^\alpha  
+ \frac{1}{4} \bigl( a_\beta \omega^\beta \bigr) a^\alpha  
\nonumber \\ & & \mbox{} 
- \frac{1}{8} \bigl( a_\beta \omega^\beta \bigr)^2 \omega^\alpha 
+ \frac{3}{8} \bigl( \dot{a}_\beta u^\beta \bigr) \omega^\alpha 
\nonumber \\ & & \mbox{} 
+ \frac{1}{3} \omega^\alpha R_{\beta\gamma\delta\epsilon} 
  \omega^\beta u^\gamma \omega^\delta u^\epsilon 
- \frac{1}{6} R^\alpha_{\ \beta\gamma\delta} 
  u^\beta \omega^\gamma u^\delta
\nonumber \\ & & \mbox{} 
+ \frac{1}{12} \bigl( 5 R_{\beta\gamma}u^\beta u^\gamma 
  + R_{\beta\gamma} \omega^\beta \omega^\gamma 
  + R \bigr) \omega^\alpha  
\nonumber \\ & & \mbox{} 
- \frac{1}{6} \bigl( \delta^\alpha_{\ \beta} + u^\alpha u_\beta \bigr)
  R^\beta_{\ \gamma} \omega^\gamma \biggr] 
+ e F^{\alpha}_{{\rm R} \beta} u^\beta 
\nonumber \\ & & \mbox{} 
+ O(s) 
\label{3.13}
\end{eqnarray}
for the force acting on the charge $e$ at position $x$, due to the
charge $\bar{e}$ moving on $\bar{\gamma}$. This expression involves
$s$, the spatial distance from $\bar{x}$ to $x$; the unit vector
$\omega^\alpha$ defined at the end of Sec.~II; the velocity vector
$u^\alpha$ of the charge $e$, its acceleration vector $a^\alpha$, and
its covariant derivative $\dot{a}^\alpha$; the Riemann and Ricci
tensors, and the Ricci scalar, all evaluated at $x$. And it involves  
the Detweiler-Whiting regular field, also evaluated at $x$ 
[Eqs.~(\ref{3.5})--(\ref{3.8}); LRR Eqs.~(473) and (474)],     
\begin{eqnarray} 
e F^{\alpha}_{{\rm R} \beta} u^\beta &=& e \bar{e} \biggl[  
\bigl( \delta^\alpha_{\ \beta} + u^\alpha u_\beta \bigr)
\biggl( \frac{2}{3} \dot{a}^{\beta} 
+ \frac{1}{3} R^{\beta}_{\ \gamma} u^{\gamma} \biggr)  
\nonumber \\ & & \hspace*{-40pt} \mbox{} 
+ 2 u_\beta \int_{-\infty}^{\bar{\tau}^-} \nabla^{[\alpha} 
G^{\beta]}_{\ \gamma'}(x,z') u^{\gamma'}\, d\bar{\tau}' + O(s)
\biggr].   
\label{3.14}
\end{eqnarray} 
It is important to note that our final expression for $f^\alpha$ is a
legitimate tensorial relation that is valid in all coordinate
systems; all traces of the Fermi normal coordinates, which were
heavily involved in the calculations leading to Eq.~(\ref{3.13}), have
been eliminated. 

\section{Dumbbell} 

We construct an extended charged body, a dumbbell, by placing a
charge $e_-$ on a accelerated world line $\gamma_-$, and by placing a
second charge $e_+$ on a neighboring world line $\gamma_+$. The
magnitude of each charge is arbitrary, but we assume that their sum
$e = e_- + e_+$, the total charge of the extended body, does not
vanish. We imagine that each charge follows a predetermined flight
plan, and we wish to determine the forces acting on each charge. In a  
suitable sense to be spelled out in the following section, the sum
total of these forces will be the net force acting on the extended
body. 

We imagine that each charge is equipped with a control system that   
consists of a sensor capable of detecting any deviation from the
charge's prescribed motion, and a propulsion system capable of
applying the necessary corrections to return the charge to its
prescribed world line.\footnote{Here we deviate from the Ori-Rosenthal 
  treatment, in which the charges are maintained in place by a rigid
  rod instead of a propulsion system. We use the artifice of a
  propulsion system to avoid the artifice of a perfectly rigid rod
  (which would violate relativistic causality) or the complications
  (such as energy dissipation) associated with a physically realistic
  rod. We do not think that this difference in modeling is very
  important; in flat spacetime the two models produce identical
  answers for the self-force.}      
The acceleration of each world line is produced by an external force
$f^\alpha_{\rm ext}$ which varies smoothly from $\gamma_-$ to
$\gamma_+$; the acceleration of each world line is related to the
external force by $m a^\alpha = f^\alpha_{\rm ext}$, where $m$ is the
mass of each charge. The role of the propulsion system is to apply a 
force $f^\alpha_{\rm prop}$ that compensates for the electromagnetic
forces produced within the dumbbell, which tend to push each charge
away from its prescribed motion. The sum total of the propulsion
forces is equal and opposite to the sum total of the forces produced
within the dumbbell.        

The motion of each charge will be referred to a reference world line
$\bar{\gamma}$ which defines the spatial origin of a Fermi coordinate 
system $(t,x^a)$. The reference world line lies between the charges,
and this new Fermi system {\it is distinct} from the one used in the
preceding section. The previous system of FNC was centered on one of
the two charges, the one that was labeled $\bar{e}$; here the FNC are
centered instead on the central world line 
$\bar{\gamma}$. Nevertheless, our construction will allow us to 
continue to use the language and results of Sec.~II. 

Because our charges are made to follow a predetermined flight plan, we
are free to choose their world lines $\gamma_\pm$. We pick two members
of the congruence of comoving world lines, and we pick them to be on
opposite sides of the reference world line; this translates into the
selection $\gamma_\pm = \gamma(x^a_\pm)$, where $x^a_\pm = \pm
\varepsilon n^a$. Here, $\varepsilon$ is the constant spatial
separation between each world line and $\bar{\gamma}$, and $n^a$ is a 
unit vector that specifies the orientation of the dumbbell. The
charges are therefore maintained at a constant spatial separation
$s = 2\varepsilon$. 

The components of each charge's velocity vector in FNC can be obtained
from Eqs.~(\ref{2.24}), (\ref{2.25}) by substituting $x^a = x^a_\pm =
\pm \varepsilon n^a$. This gives 
\begin{eqnarray} 
u^t_\pm &=& 1 \mp \varepsilon \bar{a}_a n^a 
+ \varepsilon^2 (\bar{a}_a n^a)^2  
- \frac{1}{2} \varepsilon^2 \bar{R}_{ctdt} n^c n^d 
\nonumber \\ & & \mbox{} + O(\varepsilon^3), 
\label{4.1} \\ 
u^a_\pm &=& 0, 
\label{4.2}
\end{eqnarray} 
where, as before, $\bar{a}_a$ are the components of $\bar{\gamma}$'s
acceleration vector, and $\bar{R}_{ctdt}$ are components of the
Riemann tensor evaluated on $\bar{\gamma}$. The acceleration vectors
are given by Eq.~(\ref{2.28}), 
\begin{eqnarray} 
a^t_\pm &=& O(\varepsilon^2),
\label{4.3} \\ 
a^a_\pm &=& (1 \mp \varepsilon \bar{a}_b n^b) \bar{a}^a 
\pm \varepsilon \bar{R}^a_{\ tct} n^c + O(\varepsilon^2). 
\label{4.4}
\end{eqnarray}  
 
The force $f^\alpha_+$ acting on the charge $e_+$, due to the charge
$e_-$, can be obtained from Eq.~(\ref{3.13}) by making the
substitutions $e = e_+$, $\bar{e} = e_-$, $s = 2\varepsilon$,
$\omega^t = 0$, $\omega^a = +n^a$, $u^\alpha = u^\alpha_+$, $a^\alpha
= a^\alpha_+$, and by involving
Eqs.~(\ref{4.1})--(\ref{4.4}). Similarly, the force $f^\alpha_-$ 
acting on $e_-$, due to $e_+$, can be obtained by making the
substitutions $e = e_-$, $\bar{e} = e_+$, $s = 2\varepsilon$,
$\omega^t = 0$, $\omega^a = -n^a$, $u^\alpha = u^\alpha_-$, and
$a^\alpha = a^\alpha_-$. Straightforward manipulations reveal that the
spatial components of the forces $f^\alpha_\pm$ in FNC are 
\begin{eqnarray} 
f^a_\pm &=& e_+ e_- \biggl[ 
\pm \frac{1}{4 \varepsilon^2} n^a 
- \frac{1}{4\varepsilon} \bar{a}^a
- \frac{1}{4\varepsilon} \bigl( \bar{a}_b n^b \bigr) n^a 
\nonumber \\ & & \mbox{} 
\pm \frac{1}{2} \bigl( \bar{a}_b n^b \bigr) \bar{a}^a 
\pm \frac{1}{8} \bigl( \bar{a}_b n^b \bigr)^2 n^a 
\pm \frac{3}{8} \dot{\bar{a}}_t n^a 
\nonumber \\ & & \mbox{} 
\pm \frac{1}{12} n^a \bar{R}_{btct} n^b n^c
\mp \frac{5}{12} \bar{R}^a_{\ tbt} n^b 
\nonumber \\ & & \mbox{} 
\pm \frac{1}{12} \bigl( 5 \bar{R}_{tt} + \bar{R}_{bc} n^b n^c 
                 + \bar{R} \bigr) n^a 
\nonumber \\ & & \mbox{} 
\mp \frac{1}{6} \bar{R}^a_{\ b} n^b 
+ \bar{{\gothf}}^a_{\rm R} + O(s) \biggr].
\label{4.5}
\end{eqnarray}   
The time component of the forces is $O(s)$. In the manipulations
leading to Eq.~(\ref{4.5}) we have displaced the evaluation point of
the Riemann and Ricci tensors from $\gamma_\pm$ to the reference world
line $\bar{\gamma}$. This can be done carefully by going through the
procedure outlined in the paragraph following Eq.~(\ref{3.12}); or
this can be done more simply by noting that the displacement
introduces a correction term of order $s$ that can be absorbed into
the error term, $O(s)$.    

We have introduced the rescaled Detweiler-Whiting regular force   
${\gothf}^{\bar{\alpha}}_{\rm R}$, which is also evaluated on
$\bar{\gamma}$ in Eq.~(\ref{4.5}). This is defined by removing a
factor of the charge from the expression for 
$F^{\alpha}_{{\rm R} \beta} u^\beta$ given in Eq.~(\ref{3.14}). If,
for example, $e_+ F^{-\alpha}_{{\rm R}\ \beta} u_+^\beta$ is the
contribution to the force acting on $e_+$ that comes from $e_-$'s
regular field, we write it as $e_+ e_- {\gothf}^{\alpha}_{\rm R}$ and
displace the result from $\gamma_+$ to $\bar{\gamma}$. The rescaled
regular force evaluated on $\bar{\gamma}$ is expressed in covariant
form as 
\begin{eqnarray} 
{\gothf}^{\bar{\alpha}}_{\rm R} &=&   
\bigl( \delta^{\bar{\alpha}}_{\ \bar{\beta}} + u^{\bar{\alpha}} 
u_{\bar{\beta}} \bigr) \biggl( \frac{2}{3} \dot{a}^{\bar{\beta}}   
+ \frac{1}{3} R^{\bar{\beta}}_{\ \bar{\gamma}} u^{\bar{\gamma}}
\biggr) 
\nonumber \\ & & \mbox{} 
+ 2 u_{\bar{\beta}} \int_{-\infty}^{\bar{\tau}^-} 
\nabla^{[\bar{\alpha}} G^{\bar{\beta}]}_{\ \gamma'}(\bar{x},z') 
u^{\gamma'}\, d\bar{\tau}'. 
\label{4.6}
\end{eqnarray} 
It is independent of the charges $e_\pm$, and its spatial components
in FNC are $\bar{{\gothf}}^a_{\rm R}$, which is what appears in
Eq.~(\ref{4.5}). 

\section{Total force acting on the extended body} 

The net force acting on the dumbbell of Sec.~IV consists of eight 
separate contributions. The first is $f^a_+$, the force acting on
$e_+$ due to the field created by $e_-$; the second is $f^a_-$, the
force acting on $e_-$ due to the field created by $e_+$; the third is
$f^a_{{\rm self}+}$, the self-force acting on $e_+$ due to this
charge's own field; the fourth is $f^a_{{\rm self}-}$, the
self-force acting on $e_-$; the fifth is $f^a_{{\rm prop}+}$, the
force produced by $e_+$'s propulsion system to compensate for the
action of $f^a_+$ and $f^a_{{\rm self}+}$; the sixth is 
$f^a_{{\rm prop}-}$, the force produced by $e_-$'s propulsion system;
and the remaining two are the external forces $f^a_{\rm ext}$ that are 
responsible for the acceleration of each charge.   

We recall from Sec.~I the fundamental axiom behind our derivation: 
{\it The self-force acting on a point charge is a well-defined vector
field on the particle's world line, and it is proportional to the
square of the particle's electric charge.} We will invoke these
properties in the following manipulations, but we state now that
according to the axiom, the individual self-forces 
$f^a_{{\rm self}\pm}$ {\it must be included} in the accounting of 
the net force; it would be inconsistent to leave them out, as we 
expect the extended body to retain a nonvanishing self-force in the
limit $\varepsilon \to 0$.  

In a sense to be specified below, the net force acting on the dumbbell
must be the sum total of the eight individual contributions listed
above. We denote by $f^{\bar{\alpha}}_{\rm net, int}$ the net internal 
electromagnetic force acting on the dumbbell; this is the sum total of
$f^a_+$, $f^a_-$, $f^a_{{\rm self}+}$, and $f^a_{{\rm self}-}$
evaluated on the central world line. We denote by 
$f^{\bar{\alpha}}_{\rm net, prop}$ the net force produced by the
propulsion system of each charge; this is the sum total of 
$f^a_{{\rm prop}+}$ and $f^a_{{\rm prop}-}$ evaluated on the central
world line. Finally, we denote by $f^{\bar{\alpha}}_{\rm ext}$ the net 
external force acting on the dumbbell; this is the sum total of the 
external forces acting on each charge.     

The propulsion forces are designed to compensate for the action of the
internal electromagnetic forces. We therefore have 
$f^{\bar{\alpha}}_{\rm net, prop} + f^{\bar{\alpha}}_{\rm net, int} = 
0$, and the equation of motion for the dumbbell is $m a^{\alpha} =
f^{\alpha}_{\rm ext}$, where $m$ is the total mass of the dumbbell;
the external forces are keeping the dumbbell on its prescribed world
line $\bar{\gamma}$. If we imagine ourselves momentarily switching off 
the propulsion system, the internal electromagnetic forces would no
longer be compensated for, and we would expect that the dumbbell would
start moving according to the new equation $m a^{\alpha} = 
f^{\alpha}_{\rm ext} + f^{\alpha}_{\rm self}$, with the second term
denoting the self-force acting on the dumbbell.\footnote{This equation
  is meant to apply only at the instant at which the propulsion system
  is switched off. We need not be concerned with the fact that the
  dumbbell has momentarily lost its cohesion and is about to fly 
  apart.}  
It is therefore clear that the propulsion forces are in fact
compensating for the self-force, so that  
$f^{\bar{\alpha}}_{\rm net, prop} 
= -f^{\bar{\alpha}}_{\rm self}$. This allows us to conclude that the  
self-force acting on the dumbbell is equal to 
$f^{\bar{\alpha}}_{\rm net, int}$, and this is what we shall calculate 
in the remainder of this section.     

The individual forces that make up $f^{\bar{\alpha}}_{\rm net, int}$  
cannot simply be added up, because they are vectors that are defined
at different points in the manifold of a curved spacetime. Before
taking the sum they must all be transported to a single common point,
which we choose to be $\bar{x}$ on the reference world line. We need a 
{\it transport rule} to carry out this operation.  

The correct transport rule in flat spacetime was identified by Ori and 
Rosenthal \cite{ori-rosenthal:03, ori-rosenthal:04}. Building on
considerations of momentum conservation, they showed that the
transport rule involves parallel transport of the force from
$\gamma_\pm$ to $\bar{\gamma}$ combined with multiplication by a
kinematic factor $d\tau_\pm/d\bar{\tau}$ that converts increments
of proper time on $\gamma_\pm$ to an increment of proper time on the
central world line. The parallel-transport component of the rule is
easy enough to understand, and the kinematic conversion is easily
justified by the fact that an increment of momentum on each world line
is related to the force by $dp^\alpha_\pm = f^\alpha_\pm\, d\tau_\pm$; 
because a similar statement applies to $\bar{\gamma}$, a conversion 
from $d\tau_\pm$ to $d\bar{\tau}$ is clearly required.  

We write the net internal electromagnetic force acting on the dumbbell 
as  
\begin{equation} 
f_{\rm net, int}^{\bar{\alpha}} = k^{\bar{\alpha}}_{\ \alpha}  
\bigl( f^\alpha_+ + f^\alpha_{{\rm self}+} \bigr) 
+ k^{\bar{\alpha}}_{\ \alpha} 
\bigl( f^\alpha_- + f^\alpha_{{\rm self}-} \bigr), 
\label{5.1}
\end{equation} 
where $k^{\bar{\alpha}}_{\ \alpha}(\bar{x},x)$ is the transport 
operator which takes a force at $x_\pm$ on $\gamma_\pm$ and brings it
to the point $\bar{x}$ on $\bar{\gamma}$. According to Ori and
Rosenthal, this is $k^{\bar{\alpha}}_{\ \alpha} =
(d\tau_\pm/d\bar{\tau}) g^{\bar{\alpha}}_{\ \alpha}$ in flat
spacetime, with the second factor denoting parallel transport from 
$x_\pm$ to $\bar{x}$. According to Eqs.~(\ref{2.22}) and (\ref{2.23}),
from which we remove all terms that involve the Riemann tensor, the
relevant components of the transport operator in FNC are   
\begin{equation} 
k^a_{\ b} = \bigl( 1 \pm \varepsilon \bar{a}_c n^c \bigr) 
\delta^a_{\ b} + O(\varepsilon^3) \qquad \mbox{(flat spacetime)}.   
\label{5.2}
\end{equation} 

In curved spacetime we would expect this expression to be modified by
a term of order $\varepsilon^2$ which, presumably, would be
constructed from the Riemann tensor. We shall, however, remain
uncommitted as to the exact nature of the transport rule in curved 
spacetime. We shall not assume that the flat-spacetime prescription  
$k^{\bar{\alpha}}_{\ \alpha} = (d\tau_\pm/d\bar{\tau})  
g^{\bar{\alpha}}_{\ \alpha}$ is preserved in curved spacetime. And we
shall not adopt any specific alternative prescription such as, for
example, the Dixon-Bailey-Israel transport rule \cite{dixon:70a,
dixon:70b, dixon:74, dixon:79, bailey-israel:80} that is widely used
in the description of extended bodies in general relativity. Instead,
we shall adopt a very generic transport rule of the form  
\begin{equation}
k^a_{\ b} = \bigl( 1 \pm \varepsilon \bar{a}_c n^c \bigr) 
\delta^a_{\ b} + \varepsilon^2 \bar{A}^a_{\ bcd} n^c n^d 
+ O(\varepsilon^3), 
\label{5.3}
\end{equation} 
with $\bar{A}^a_{\ bcd}$ representing the components of an unknown
tensor which is evaluated at $\bar{x}$. We may imagine that this
tensor is formed from the Riemann tensor, but we do not need to
make such a statement; our only requirement here is that the
quantities $\bar{A}^a_{\ bcd}$ must depend on $\bar{x}$ only and be 
independent of $x_\pm$. The only allowed dependence of $k^a_{\ b}$ on
$x_\pm^a = \pm\varepsilon n^a$ is the one which is shown explicitly in 
Eq.~(\ref{5.3}). 

Observe that Eq.~(\ref{5.3}) can be viewed as representing the Taylor
series of a general bitensor $k^a_{\ b}$ in powers of $x^a$. The first
two terms in the series are determined by the requirement that 
$k^a_{\ b}$ must reduce to the expression of Eq.~(\ref{5.2}) in flat
spacetime. The term of order $\varepsilon^2$, however, is
arbitrary. Its only relevant property, for the purposes of our
calculation, is that {\it it comes with the same sign} whether the
transport originates from $x_-$ or $x_+$; unlike the term of order
$\varepsilon$, it does not alternate in sign.  

Combining Eqs.~(\ref{4.5}) and (\ref{5.3}) gives 
\begin{eqnarray} 
k^a_{\ b} f^b_\pm &=& e_+ e_- \biggl[ 
\pm \frac{1}{4 \varepsilon^2} n^a 
- \frac{1}{4\varepsilon} \bar{a}^a
\pm \frac{1}{4} \bigl( \bar{a}_b n^b \bigr) \bar{a}^a 
\nonumber \\ & & \mbox{} 
\mp \frac{1}{8} \bigl( \bar{a}_b n^b \bigr)^2 n^a 
\pm \frac{3}{8} \dot{\bar{a}}_t n^a 
\pm \frac{1}{12} n^a \bar{R}_{btct} n^b n^c
\nonumber \\ & & \mbox{} 
\mp \frac{5}{12} \bar{R}^a_{\ tbt} n^b 
\pm \frac{1}{12} \bigl( 5 \bar{R}_{tt} + \bar{R}_{bc} n^b n^c 
                 + \bar{R} \bigr) n^a 
\nonumber \\ & & \mbox{} 
\mp \frac{1}{6} \bar{R}^a_{\ b} n^b
\pm \frac{1}{4} \bar{A}^a_{\ bcd} n^b n^c n^d  
\nonumber \\ & & \mbox{} 
+ \bar{{\gothf}}^a_{\rm R} + O(\varepsilon) \biggr],      
\label{5.4}
\end{eqnarray}   
and from this it follows that 
\begin{equation} 
k^a_{\ b} f^b_+ + k^a_{\ b} f^b_- = e_+ e_- \biggr[ 
- \frac{1}{2\varepsilon} \bar{a}^a
+ 2\bar{{\gothf}}^a_{\rm R} + O(\varepsilon) \biggr]. 
\label{5.5}
\end{equation} 
This result will be substituted into Eq.~(\ref{5.1}) in a moment. 

The net internal electromagnetic force acting on the dumbbell involves
also the combined action of the individual self-forces 
$f^a_{{\rm self}\pm}$. Here we invoke our axiom and write  
\begin{equation} 
f^a_{{\rm self}\pm} = e_\pm^2 {\gothf}^a_{\rm self}[\gamma_\pm],  
\label{5.6} 
\end{equation} 
which makes the statement that each self-force is proportional to the
square of the particle's charge; the rescaled self-force 
${\gothf}^a_{\rm self}[\gamma_\pm]$ is independent of $e_\pm$. As
indicated, each rescaled self-force is a functional of a world line
which, we presume, can be defined in terms of the velocity vector and
its derivatives, and in terms of spacetime tensors (and bitensors)
that are evaluated on the world line. We assume that this is 
{\it the same functional} for each world line, and that the result is
a smooth vector field on the world line. From Eqs.~(\ref{5.3}) and
(\ref{5.6}) we get  
\begin{equation} 
k^a_{\ b} f^b_{{\rm self}+} + k^a_{\ b} f^b_{{\rm self}-} = 
(e_+^2 + e_-^2) \bar{{\gothf}}^a_{\rm self}[\bar{\gamma}] 
+ O(\varepsilon), 
\label{5.7}
\end{equation} 
in which the rescaled self-force is now evaluated on the central world
line $\bar{\gamma}$. 

Equation (\ref{5.1}) now reads 
\begin{eqnarray}  
\bar{f}^a_{\rm net, int} &=& -\frac{e_+ e_-}{s} \bar{a}^a  
+ 2 e_+ e_- \bar{{\gothf}}^a_{\rm R} 
+ (e_+^2 + e_-^2) \bar{{\gothf}}^a_{\rm self}[\bar{\gamma}] 
\nonumber \\ & & \mbox{} + O(s),  
\label{5.8} 
\end{eqnarray}  
where $s = 2\varepsilon$ is the spatial separation between the two
charges. Equation (\ref{5.8}) gives the components of the net force in
FNC.  

The first term on the right-hand side of Eq.~(\ref{5.8}) is divergent
in the limit $s \to 0$. It has the form of $-m_{\rm el} a^a$, where  
\begin{equation} 
m_{\rm el} = \frac{e_+ e_-}{s} 
\label{5.9}
\end{equation} 
is the electromagnetic contribution --- its electrostatic energy ---
to the mass of the dumbbell. This term produces an additional inertial
term in the equations of motion, and it can be combined with the term
$m_{\rm material} a^a$ to form the appropriate left-hand side,
$m_{\rm inertial} a^a = (m_{\rm material} + m_{\rm el}) a^a$. Here  
$m_{\rm material}$ is the purely material contribution to the
dumbbell's mass, to which the electromagnetic contribution 
$m_{\rm el}$ is added to form $m_{\rm inertial}$, the dumbbell's total
inertial mass. The diverging term has therefore been eliminated by
mass renormalization, and at this stage there is no obstacle in taking
the limit $s \to 0$ of Eq.~(\ref{5.8}); the dumbbell finally becomes a 
point particle. 

As was first pointed out by Ori and Rosenthal \cite{ori-rosenthal:03,
ori-rosenthal:04}, the mass-renormalization procedure is successful
because the diverging term in $\bar{f}_{\rm net}^a$ is precisely
proportional to the acceleration $\bar{a}^a$; this is a true property
of the net force because the directional terms $(\bar{a}_b n^b)
n^a/(4\varepsilon)$ that appeared in the mutual forces $f^a_\pm$ were 
eliminated during the transport from $\gamma_\pm$ to
$\bar{\gamma}$. As they also pointed out, the procedure is successful
also because it is not subjected to the famous 4/3 problem
\cite{rohrlich:90}: the shift in the mass is $m_{\rm el}$ and not 
$\frac{4}{3} m_{\rm el}$.   

The second and third terms in Eq.~(\ref{5.8}) combine to form the
dumbbell's own self-force, 
\begin{equation} 
\bar{f}^a_{\rm self} = 2 e_+ e_- \bar{{\gothf}}^a_{\rm R} 
+ (e_+^2 + e_-^2) \bar{{\gothf}}^a_{\rm self}[\bar{\gamma}]. 
\label{5.10}
\end{equation} 
We once more invoke our axiom and write the self-force as 
\begin{equation} 
\bar{f}^a_{\rm self} = e^2 \bar{\gothf}^a_{\rm self}[\bar{\gamma}],  
\label{5.11}
\end{equation}
in terms of the total charge $e = e_+ + e_-$ and in terms of the same
rescaled self-force functional that was introduced in
Eq.~(\ref{5.6}). Equating Eqs.~(\ref{5.10}) and (\ref{5.11}) produces
an equation to be solved for 
$\bar{{\gothf}}^a_{\rm self}[\bar{\gamma}]$, and we obtain our central 
result,   
\begin{equation}
\bar{{\gothf}}^a_{\rm self}[\bar{\gamma}] = \bar{{\gothf}}^a_{\rm R}.  
\label{5.12}
\end{equation} 
The rescaled self-force is the Detweiler-Whiting rescaled regular
force of Eq.~(\ref{4.6}). 

Our final conclusion is that the self-force acting on a point particle
of charge $e$ is given by  
\begin{eqnarray} 
f^{\bar{\alpha}}_{\rm self} &=& e^2 
\bigl( \delta^{\bar{\alpha}}_{\ \bar{\beta}} + u^{\bar{\alpha}}  
u_{\bar{\beta}} \bigr) \biggl( \frac{2}{3} \dot{a}^{\bar{\beta}}   
+ \frac{1}{3} R^{\bar{\beta}}_{\ \bar{\gamma}} u^{\bar{\gamma}}
\biggr) 
\nonumber \\ & & \mbox{} 
+ 2 e^2 u_{\bar{\beta}} \int_{-\infty}^{\bar{\tau}^-} 
\nabla^{[\bar{\alpha}} G^{\bar{\beta}]}_{\ \gamma'}(\bar{x},z') 
u^{\gamma'}\, d\bar{\tau}'. 
\label{5.13}
\end{eqnarray} 
This is the standard expression for the self-force. The particle's
equations of motion are then 
\begin{equation} 
m a^{\bar{\alpha}} = f^{\bar{\alpha}}_{\rm self}
+ f^{\bar{\alpha}}_{\rm ext}, 
\label{5.14}
\end{equation}  
with the inertial mass $m$ identified with the combined material and
electromagnetic mass of the charged body. 

\section{Discussion: Critical review of existing derivations of the
self-force}  

We review in this section the various approaches that have been used
in the literature to derive the standard expression for the
electromagnetic self-force in curved spacetime, as given by
Eq.~(\ref{1.2}). While all these approaches have led to the same
answer, they possess different strengths and weaknesses, and none of
them are completely immune to criticism. This critical review 
motivated our own approach to the problem, which is also summarized at
the end of this section. The presentation in this section borrows
heavily from [LRR Sec.~5.5].  

\subsection{Conservation of energy-momentum} 

Conservation of energy-momentum was used by Dirac \cite{dirac:38} to
derive the equations of motion of a point electric charge in flat
spacetime, and the same method was adopted by DeWitt and Brehme
\cite{dewitt-brehme:60} in their generalization of Dirac's work to
curved spacetime. This approach, of course, produced the first
derivation of the standard expression. (The DeWitt-Brehme derivation
contained a calculational mistake that was later corrected by 
Hobbs \cite{hobbs:68}.)      

The method is based on the conservation equation 
$T^{\alpha\beta}_{\ \ \ ;\beta} = 0$, where the stress-energy tensor
$T^{\alpha\beta}$ includes a contribution from the charged particle
and a contribution from the electromagnetic field. The particle's
contribution is a Dirac distribution on the world line, and the
field's contribution diverges as $1/s^4$ near the world line, where
$s$ is the spatial distance from the field point to the world
line. (We are using Fermi normal coordinates in this discussion.)
While in flat spacetime the differential statement of energy-momentum
conservation can immediately be turned into an integral statement, the
same is not true in a curved spacetime (unless the spacetime possesses
at least one Killing vector). To proceed it is necessary to rewrite
the conservation equation as   
\begin{equation} 
0 = g^{\bar{\mu}}_{\ \alpha} T^{\alpha\beta}_{\ \ \ ;\beta}  
= \bigl( g^{\bar{\mu}}_{\ \alpha} T^{\alpha\beta} \bigr)_{;\beta}  
- g^{\bar{\mu}}_{\ \alpha;\beta} T^{\alpha\beta}, 
\label{6.1}
\end{equation} 
where $g^{\bar{\mu}}_{\ \alpha}(\bar{x},x)$ is the parallel propagator
from $x$ to an arbitrary point $\bar{x}$ on the world
line. Integrating this equation over the interior of a world-tube
segment that consists of a ``wall'' of constant $s$ and two ``caps''
of constant $t$, we obtain    
\begin{eqnarray} 
0 &=& 
\int_{\rm wall} g^{\bar{\mu}}_{\ \alpha} T^{\alpha\beta} d\Sigma_\beta  
+ \int_{\rm caps} g^{\bar{\mu}}_{\ \alpha} T^{\alpha\beta} d\Sigma_\beta  
\nonumber \\ & & \mbox{} 
+ \int_{\rm interior} g^{\bar{\mu}}_{\ \alpha;\beta} T^{\alpha\beta}\,
dV,  
\label{6.2}
\end{eqnarray} 
where $d\Sigma_\beta$ is a three-dimensional surface element and $dV$
an invariant, four-dimensional volume element. 

There is no obstacle in evaluating the wall integral, for which
$T^{\alpha\beta}$ reduces to the field's stress-energy tensor; for a
wall of radius $s$ the integral scales as $1/s^2$. The integrations 
over the caps, however, are problematic: While the particle's
contribution to the stress-energy tensor is integrable, the
integration over the field's contribution goes as $\int_0^s
(s')^{-2}\, ds'$ and diverges. This issue arises also in flat
spacetime \cite{dirac:38}, and in this case the regularization of the
cap integrations, and the removal of all singular terms by mass
renormalization, has been done with great rigor using robust
distributional methods. The state of the art for flat spacetime was
reviewed in a paper by Teitelboim, Villarroel, and van Weert
\cite{teitelboim-etal:80}. As far as we are aware, however, the cap 
integrations for curved spacetime were never handled with such care,
and the DeWitt-Brehme derivation of the self-force does not enjoy the
same degree of rigor as what has been achieved for flat spacetime.    

More troublesome, however, is the treatment of the interior integral,
which does not occur in flat spacetime, and which was simply
discarded by DeWitt and Brehme. Because 
$g^{\bar{\mu}}_{\ \alpha;\beta}$ scales as $s$, the field's 
stress-energy tensor as $1/s^4$, and $dV$ as $s^2\, dsdt$, this
integral goes as $\int_0^s (s')^{-1}\, ds' dt$ and it also diverges,
albeit less strongly than the caps integration. While simply
discarding this integral does produce the standard expression for the 
electromagnetic self-force, it would be desirable to go through a
careful regularization of the interior integration, and find a
convincing reason to discard it altogether. To the best of our
knowledge, this has not been done, and until this issue is fully
resolved the DeWitt-Brehme derivation shall be open to criticism.  
              
\subsection{Averaging method} 

An alternative derivation of the electromagnetic self-force is
presented in [LRR Sec.~5.2.6]. It is based on the idea that it is the 
average of the retarded field around the charged particle which is
responsible for the net force acting on it. This derivation has the
advantage of being much easier to implement than a derivation based on
energy-momentum conservation; it requires a computation of the
retarded field of a point charge, but it does not require the
computation of the field's stress-energy tensor. 

To describe the averaging method it is convenient to adopt the
Detweiler-Whiting decomposition of the retarded field
$F_{\alpha\beta}$ into singular and regular pieces   
\cite{detweiler-whiting:03},  
\begin{equation}
F_{\alpha\beta} = F^{\rm S}_{\alpha\beta} + F^{\rm R}_{\alpha\beta}.  
\label{6.3}
\end{equation} 
As was shown by Detweiler and Whiting, this decomposition is
unambiguous, the retarded and singular fields share the same
singularity structure near the world line, and the regular field 
$F^{\rm R}_{\alpha\beta}$ is smooth on the world line. 

To formulate equations of motion for the point charge we temporarily 
model it as a spherical hollow shell, and we obtain the net force
acting on this object by averaging $F_{\alpha\beta}$ over the shell's
surface. (The averaging is performed in the shell's rest frame,
and the shell is spherical in the sense that its proper distance $s$
from the world line is the same in all directions.) The averaged field
is next evaluated on the world line, in the limit of a zero-radius
shell. Because the regular field is smooth on the world line, this
yields   
\begin{equation} 
e\langle F_{\bar{\mu}\bar{\nu}} \rangle u^{\bar{\nu}} 
= e\langle F^{\rm S}_{\bar{\mu}\bar{\nu}} \rangle u^{\bar{\nu}} 
+ e F^{\rm R}_{\bar{\mu}\bar{\nu}} u^{\bar{\nu}},  
\label{6.4}
\end{equation} 
where $u^{\bar{\mu}}$ is the particle's velocity vector. As was
calculated in [LRR Eq.~(475)],
\begin{equation} 
e\langle F^{\rm S}_{\bar{\mu}\bar{\nu}} \rangle u^{\bar{\nu}} 
= -(m'_{\rm el}) a_{\bar{\mu}}, \qquad  
m'_{\rm el} = \frac{2}{3} \frac{e^2}{s},   
\label{6.5}
\end{equation} 
and $e F^{\rm R}_{\bar{\mu}\bar{\nu}} u^{\bar{\nu}}$ was given in 
Eq.~(\ref{5.13}). The equations of motion are then postulated to be
$m_{\rm material} a_{\bar{\mu}} = e\langle F_{\bar{\mu}\bar{\nu}}
\rangle u^{\bar{\nu}} + f^{\rm ext}_{\bar{\mu}}$, where 
$m_{\rm material}$ is the particle's material mass. With the preceding
results we arrive at $m_{\rm inertial} a_{\bar{\mu}} 
= e F^{\rm R}_{\bar{\mu}\bar{\nu}} u^{\bar{\nu}}
+ f^{\rm ext}_{\bar{\mu}}$, where $m_{\rm inertial} \equiv 
m_{\rm material} + m'_{\rm el}$ is the particle's observed
(renormalized) inertial mass. We have recovered the standard 
expression for the self-force, $f^{\rm self}_{\bar{\mu}} = 
e F^{\rm R}_{\bar{\mu}\bar{\nu}} u^{\bar{\nu}}$.    

The averaging method is basically sound, but it contains a number of
weak points. A first source of criticism concerns the specifics of the 
averaging procedure, in particular, the choice of a spherical surface
over any other conceivable shape. Another source is an inconsistency
of the method which gives rise to the famous ``4/3 problem''
\cite{rohrlich:90}: the mass shift $m'_{\rm el}$ is related to the
shell's electrostatic energy $m_{\rm el} = e^2/(2s)$ by $m'_{\rm el} = 
\frac{4}{3} m_{\rm el}$ instead of the expected $m'_{\rm el} 
= m_{\rm el}$. This problem originates from the fact that the field
that is averaged over the surface of the shell is sourced by a point
particle and not by the shell itself. It is likely that a more careful
treatment of the near-source field will eliminate both sources of
criticism: We can expect that the field produced by an extended
spherical object will give rise to a mass shift that equals the
object's electrostatic energy, and the object's spherical shape would
then fully justify a spherical averaging. (Considering other shapes
might also be possible, but one would prefer to keep the object's
structure simple and avoid introducing additional multipole moments.)
Further work is required to clean up these details.   

The averaging method is at the core of the approach followed by Quinn
and Wald \cite{quinn-wald:97}, who also average the retarded field
over a spherical surface surrounding the particle. Their approach,
however, also incorporates a ``comparison axiom'' that eliminates the 
need to explicitly renormalize the mass. Their axiomatic approach is
perfectly sound, but it completely bypasses the interesting question
of determining the electromagnetic contribution to the particle's
inertia.  

\subsection{Detweiler-Whiting axiom} 

The Detweiler-Whiting decomposition of the retarded field
\cite{detweiler-whiting:03} becomes most powerful when it is combined
with the Detweiler-Whiting axiom, which asserts that  
\begin{verse} 
the singular field $F^{\rm S}_{\alpha\beta}$ exerts no force on the
particle (it merely contributes to the particle's inertia); the entire
self-force arises from the action of the regular field 
$F^{\rm R}_{\alpha\beta}$. 
\end{verse} 
This axiom, which is motivated by the time-symmetric nature of the
singular field and its causal relation with the world line, gives rise
to the equations of motion $m a_{\bar{\mu}} 
= e F^{\rm R}_{\bar{\mu}\bar{\nu}} u^{\bar{\nu}}
+ f^{\rm ext}_{\bar{\mu}}$, in agreement with the averaging method
(but with an implicit, instead of explicit, mass shift), and in
agreement with the standard expression for the electromagnetic
self-force. In this picture, the particle simply interacts with a free
field whose origin can be traced to the particle's past, and the
procedure of mass renormalization is again sidestepped. This picture
of a particle interacting with a free field is compelling, and it
removes any tension between the nongeodesic motion of the charge and
the principle of equivalence. But the Detweiler-Whiting axiom is quite
strong, and one would like to justify it on the basis of an
independent derivation of the self-force. This observation leads us
toward an extended-body approach as described in Sec.~I, and toward
our particular partial implementation of this program.  

\subsection{Extended-body approach} 

The derivation of the electromagnetic self-force presented in this
paper is a generalization to curved spacetime of the extended-body
approach of Ori and Rosenthal \cite{ori-rosenthal:03,
ori-rosenthal:04}, which was limited to flat spacetime. It is an
attempt to eliminate the difficulties associated with a point particle  
by dealing instead with an extended charge distribution. Our attempt
is only partially successful: While we do introduce an extended body,
it still consists of two point charges, and we make no attempt to deal
with the internal dynamics, which we replace with a contrived and
artificial control system that keeps the individual charges on their
predetermined world lines. The result, to be sure, is a derivation of
the standard expression for the self-force that lacks a strong
physical backbone. But while we readily acknowledge this major flaw,
we stress nevertheless that our derivation is logically sound and
consistent, and that it leads to the correct answer displayed in
Eq.~(\ref{1.2}). Taking the view that all details regarding internal
structure must become irrelevant in the limit $s \to 0$, perhaps it is
not so objectionable that in a first attempt to describe the motion of
an extended charged body in curved spacetime, one would adopt an
internal structure that offers simplicity at the cost of realism.       

Our extended body does indeed have the simplest conceivable shape: It 
is a dumbbell consisting of two point charges maintained at a constant
spacelike separation $s$; the dumbbell's orientation relative to the
direction of motion is arbitrary. The internal electromagnetic forces
acting on this extended body can be broken down into four
contributions: The first two are the mutual forces that each charge
exerts on the other, and the remaining two are the self-forces of each
individual charge. The self-force acting on the entire extended body
is the sum total of these four contributions, combined appropriately
in a way that respects the curved nature of the spacetime. To identify
the proper way of adding the forces was a key aspect of our
derivation; here we followed the Ori-Rosenthal prescription and
generalized it to curved spacetime. 

Another key aspect of this derivation was our reliance on the fact
that {\it each point charge within the dumbbell is subjected to its
own self-force}. That this is so may seem peculiar, but it is in fact 
required by physical consistency: It would not do to assume that the
individual self-forces do not exist and to find that a nonzero
self-force eventually emerges from the extended body in the limit 
$s \to 0$. A successful extended-body calculation based on a
collection of point charges must therefore incorporate the idea that
each charge is subjected to its own self-force.  

The extended-body approach is a consistent method of derivation that
yields the standard expression for the electromagnetic self-force. It
does, however, require an input axiom, which we reproduce from Sec.~I:  
\begin{verse} 
The self-force acting on a point charge is a well-defined vector field
on the particle's world line, and it is proportional to the square of
the particle's electric charge.
\end{verse} 
The fact that the self-force must exist and be a well-defined quantity
was motivated in the preceding paragraph. The other important aspect
of the axiom is the statement that the self-force must scale like the
square of the charge. That this must be so follows again from a
consistency requirement: In a spacetime with a timelike Killing
vector, the total work done by the self-force must equal the total
energy radiated by the particle \cite{quinn-wald:99}; because the
radiated energy is proportional to the square of the charge, the same
should be required of the self-force. 

As was discussed in Sec.~I, the need for such an axiom is unfortunate,
and we must add this flaw to the flaws that have already been
identified. Such an axiom would not be required in an extended-body
calculation that would deal honestly with the body's internal
dynamics. This flaw, therefore, is very much a consequence of our
refusal to introduce a physical model for the extended charge
distribution and the cohesive forces that hold it together. In  
addition, the assumed equality between radiated energy and work done 
by the self-force must be only approximately valid, because heat 
dissipation within an extended body would also contribute to the
energy balance. This effect, however, should become unimportant in the 
limit $s \to 0$, that is, in the limit in which the internal dynamics
decouple from the external dynamics.       

We regard this axiom as a necessary evil, but we see it also as a
fairly mild assumption. It is, in particular, milder than the
Detweiler-Whiting axiom, and we may view the extended-body approach as
eventually providing a means to relax the main assumption behind the  
Detweiler-Whiting approach. What we would have, then, is an
independent justification of the statement that the singular field
exerts no force on the particle, but that it makes an
electrostatic-energy contribution to its inertial mass.  

We have tried to present a fair and balanced assessment of the
weaknesses and strengths of our dumbbell model. While our derivation
of the electromagnetic self-force does produce the standard expression
of Eq.~(\ref{1.2}), it leaves much to be desired in terms of physical
realism. Our refusal to deal with the body's internal dynamics comes
with a high price, and our model is indeed too artificial and
contrived (though still consistent) to be fully satisfactory. It would
be desirable to return to this subject with a more physical model for
the extended charge distribution. We believe that the calculational
tools introduced in this paper will be useful in this endeavor. In
particular, it is clear to us that the problem is best formulated in
terms of Fermi normal coordinates, which incorporate in a general and
straightforward way the curved nature of the spacetime. Another
valuable contribution is our demonstration, in Sec.~V, that the
addition of partial forces acting within the extended body is largely 
insensitive to the details of the transport rule that carries these
forces to a common point. We hope that in the future, these techniques
will facilitate the formulation of a more realistic internal model.        

\acknowledgments 

We thank Eran Rosenthal for useful discussions. We also thank an
anonymous referee for vary valuable comments on an earlier version of
this article, in which the flaws of our approach were not so clearly
identified. This work was supported by the Natural Sciences and
Engineering Research Council of Canada.  

\bibliography{../bib/master} 
\end{document}